\documentclass[preprint,twocolumn]{elsarticle}
\pdfoutput=1

\usepackage[T1]{fontenc}

\usepackage[margin=2.25cm]{geometry}

\usepackage{amsmath}
\usepackage{amssymb}
\usepackage{amsthm}

\usepackage{thmtools}

\usepackage{enumitem}

\usepackage{algorithmic}
\usepackage[linesnumbered,titlenumbered,ruled,vlined,resetcount,algosection]{algorithm2e}
\SetNlSty{textbf}{\color{black}}{}

\SetCommentSty{mycommentfont}

\newcommand{\KwInTwo}[1] {\textbf{In:} #1\\}
\newcommand{\KwOutTwo}[1] {\textbf{Out:} #1\\}

\usepackage[usenames,dvipsnames]{xcolor}
\usepackage{longtable}
\usepackage{booktabs}
\usepackage{graphicx}
\usepackage{multirow}
\usepackage{float}
\usepackage{color}

\usepackage{xparse} 
\usepackage{xifthen} 

\usepackage[labelformat=simple]{subcaption}

\usepackage{todonotes}

\usepackage[hidelinks]{hyperref}
\hypersetup{
	colorlinks,
	linkcolor={black!100!black},
	citecolor={blue!100!black},
	urlcolor={blue!50!black}
}

\usepackage{pdflscape}

\usepackage[normalem]{ulem}

\usepackage{etoolbox}

\definecolor{JLcolor}{rgb}{0.740, 0.444, 0.000}

\newcommand \bigO[1] {O(#1)}

\newcommand \Root {r}
\newcommand \Pos {p}

\NewDocumentCommand \Ftree { O{T} }{ #1 }
\NewDocumentCommand \Rtree { O{\Root} O{\Ftree} }{ #2^{#1} }
\NewDocumentCommand \SubRtree { m O{\Root} O{\Ftree} }{ #3_{#1}^{#2} }
\NewDocumentCommand \SortedSubRtree { m O{} O{\Root} O{\Ftree} }{
	\ifthenelse{\isempty{#2}}
    {#4_{#1}^{#3}}
    {#4_{#2,#1}^{#3}}
}

\NewDocumentCommand \Nvert { m } { n_{#1} }

\newcommand \degree[1] {d_{#1}}

\NewDocumentCommand \longdeg { m O{\Rtree} } { d_{#2}(#1) }

\newcommand \neighs[1] {\Gamma_{#1}}

\NewDocumentCommand \longneighs { m O{\Rtree} } { \Gamma_{#2}(#1) }

\NewDocumentCommand \arr { } {\pi}

\NewDocumentCommand \Proj { O{\Rtree} }{ \mathbf{P_r}(#1)}
\NewDocumentCommand \NProj { O{\Rtree} }{ \mathbf{N_r}(#1)}
\NewDocumentCommand \DProj { O{\Rtree} }{ \mathbf{D_r}(#1)}
\NewDocumentCommand \ExpeDProj { O{\Rtree} }{\rexpe{D(#1)}}
\NewDocumentCommand \ExpeDPlan { O{\Ftree} }{\lexpe{D(#1)}}
\NewDocumentCommand \VarDProj { O{\Rtree} }{\rvar{D(#1)}}
\NewDocumentCommand \ExpeDUnc { O{\Ftree} }{\expe{D(#1)}}
\NewDocumentCommand \ZExpeDUnc { O{\Ftree} }{\Zexpe{D(#1)}}
\NewDocumentCommand \VarDUnc { O{\Ftree} }{\var{D(#1)}}

\NewDocumentCommand \PProj { O{\Pos} O{\Rtree} }{ \mathbf{P_r}(#2;#1) }
\NewDocumentCommand \PNProj { O{\Pos} O{\Rtree} }{ \mathbf{N_r}(#2;#1) }

\journal{Information Processing Letters}

\begin{document}

\twocolumn[{
\begin{frontmatter}
\title{Minimum projective linearizations of trees in linear time}

\author[affiliation1]{Llu\'is Alemany-Puig\corref{cor1}}
\ead{lalemany@cs.upc.edu}
\cortext[cor1]{Corresponding author}

\author[affiliation2]{Juan Luis Esteban}
\ead{esteban@cs.upc.edu}

\author[affiliation1]{Ramon Ferrer-i-Cancho}
\ead{rferrer@cs.upc.edu}

\affiliation[affiliation1]{
	organization={Complexity and Quantitative Linguistics Lab, LARCA Research Group, Computer Science Department, Universitat Polit\`ecnica de Catalunya},
	addressline={Jordi Girona 1-3},
	postcode={08034},
	city={Barcelona},
	country={Spain}
}

\affiliation[affiliation2]{
	organization={Computer Science Department, Universitat Polit\`ecnica de Catalunya},
	addressline={Jordi Girona 1-3},
	postcode={08034},
	city={Barcelona},
	country={Spain}
}

\begin{abstract}
The Minimum Linear Arrangement problem (MLA) consists of finding a mapping $\arr$ from vertices of a graph to distinct integers that minimizes $\sum_{\{u,v\}\in E}|\arr(u) - \arr(v)|$. In that setting, vertices are often assumed to lie on a horizontal line and edges are drawn as semicircles above said line. For trees, various algorithms are available to solve the problem in polynomial time in $n=|V|$. There exist variants of the MLA in which the arrangements are constrained. Iordanskii, and later Hochberg and Stallmann (HS), put forward $\bigO{n}$-time algorithms that solve the problem when arrangements are constrained to be planar (also known as one-page book embeddings). We also consider linear arrangements of rooted trees that are constrained to be projective (planar embeddings where the root is not covered by any edge). Gildea and Temperley (GT) sketched an algorithm for projective arrangements which they claimed runs in $\bigO{n}$ but did not provide any justification of its cost. In contrast, Park and Levy claimed that GT's algorithm runs in $\bigO{n \log d_{max}}$ where $d_{max}$ is the maximum degree but did not provide sufficient detail. Here we correct an error in HS's algorithm for the planar case, show its relationship with the projective case, and derive simple algorithms for the projective and planar cases that run without a doubt in $\bigO{n}$ time.
\end{abstract}


\end{frontmatter}
}]

\section{Introduction}
\label{sec:introduction}

A linear arrangement $\arr$ of a graph $G=(V,E)$ is a linear ordering of its vertices (it can also be seen as a permutation), i.e., vertices lie on a horizontal line. In such arrangement, the distance $d(u,v)$ between two vertices $u,v$ can be defined as $d(u,v)=|\arr(u) - \arr(v)|$ where $\arr$ maps the $n$ vertices to the $n$ distinct integers in $[1,n]$. The minimum linear arrangement problem (MLA) consists of finding a $\arr$ that minimizes the cost $D=\sum_{\{u,v\}\in E}d(u,v)$ \cite{Garey1976a,Chung1984a}. In arbitrary graphs, the problem is NP-hard \cite{Garey1976a}. For trees, various algorithms are available to solve the problem in polynomial time \cite{Goldberg1976a,Shiloach1979a,Chung1984a}. Goldberg and Klipker \cite{Goldberg1976a} devised an $\bigO{n^3}$ algorithm. Later, Shiloach \cite{Shiloach1979a} contributed with an $\bigO{n^{2.2}}$ algorithm. Finally, Chung \cite{Chung1984a} contributed with two algorithms running in $\bigO{n^2}$ time and $\bigO{n^\lambda}$ time, respectively, where $\lambda$ is any real number satisfying $\lambda>\log 3/\log 2$. The latter algorithm is the best algorithm known.

There exist several variants of the MLA problem; two of them are the {\em planar} and the {\em projective} variants. In the {\em planar} variant, namely the MLA problem under the planarity constraint, the placement of the vertices of a free tree is constrained so that there are no edge crossings. These arrangements are known as {\em planar} arrangements \cite{Kuhlmann2006a}, and also one-page book embeddings \cite{Bernhart1974a}. Two undirected edges of a graph $\{s,t\},\{u,v\}\in E$ cross if $\arr(s) < \arr(u) < \arr(t) < \arr(v)$ when, without loss of generality, $\arr(s)<\arr(t)$, $\arr(u)<\arr(v)$ and $\arr(s)<\arr(u)$. To the best of our knowledge, the first $\bigO{n}$ algorithm was put forward by Iordanskii \cite{Iordanskii1987a}. Sixteen years later, Hochberg and Stallmann (HS) put forward another $\bigO{n}$-time algorithm \cite{Hochberg2003a}. However, their algorithm contains an error which is corrected in this paper.

\begin{figure}
	\centering
	\includegraphics[scale=1]{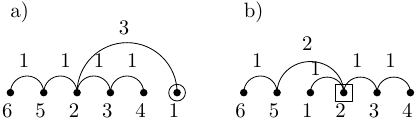}
	\caption{Two different linear arrangements of the same free tree $\Ftree$. a) A minimum projective arrangement of $\Ftree$ rooted at $1$ with cost $D=7$; the circled dot denotes the root. b) A minimum planar arrangement of $\Ftree$ with cost $D=6$ under the planarity constraint; the squared dot denotes its (only) centroidal vertex.}
	\label{fig:OPl:first_proj_gt_pla}
\end{figure}

In the {\em projective} variant, namely the MLA problem under the projectivity constraint, a rooted tree is arranged so that there are no edge crossings (i.e., the arrangement is planar) and the root is not covered. These arrangements are known as projective \cite{Kuhlmann2006a,Melcuk1988a}. A vertex $w$ is covered by an edge $\{u,v\}$ if $\arr(u)<\arr(w)<\arr(v)$ when, without loss of generality, $\arr(u)<\arr(v)$. Fig. \ref{fig:OPl:first_proj_gt_pla}(a) shows a projective arrangement while Fig. \ref{fig:OPl:first_proj_gt_pla}(b) shows an arrangement that is projective if we take vertex 2 as the root but not if we take vertex 1 as the root. Gildea and Temperley (GT) \cite{Gildea2007a} sketched an algorithm to solve this variant. The tree shown in Fig. \ref{fig:OPl:first_proj_gt_pla} is the smallest tree for which there is a vertex that, when chosen as the root, makes the minimum cost of the projective case be greater than that of the planar case (there are no other 6-vertex trees where that happens). While GT claimed that their sketch runs in $\bigO{n}$ \cite[p. 2]{Gildea2007a}, Park and Levy (PL) argued that it runs in time $\bigO{n \log d_{max}}$, where $d_{max}$ is the maximum degree. However, PL did not give enough detail to support their conclusion \cite{Park2009a}. In this article, we show that this is an overestimation of the actual complexity: the problem can be actually solved in $\bigO{n}$ time.

The remainder of the article is organized as follows. Section \ref{sec:notation_review} introduces the notation and reviews HS's algorithm. Section \ref{sec:min_planar} corrects and completes HS's algorithm \cite{Hochberg2003a}. The error is located in a recursive subprocedure ({\tt embed\_branch}) of HS's algorithm. In Section \ref{sec:min_projective}, we present two detailed $\bigO{n}$-time algorithms for the projective case that stem from HS's algorithm. HS's algorithm already contained a `subalgorithm' for solving the projective case although the authors did not identify it as such in their article \cite{Hochberg2003a}. Indeed, their algorithm can be reinterpreted as consisting of two main steps: finding a centroidal\footnote{ In this paper we follow the same terminology and notation as in \cite[Pages 35-36]{Harary1969a}. Therefore, we consider the {\em center} to be the set of {\em central} vertices, the vertices whose eccentricity is equal to the radius, and the {\em centroid} to be the set of {\em centroidal} vertices, the set of vertices whose weight, i.e., the size of the largest subtree, is minimum.} vertex (as in Iordanskii's algorithm \cite{Iordanskii1987a}) and then solving the projective case for the input tree rooted at that vertex. Hence the first algorithm for the projective case is obtained extracting the relevant part from HS's original algorithm, completing and simplifying it and, critically, using the correction indicated in Section \ref{sec:min_planar}. Our second algorithm for the projective case is a re-engineered version based on intervals that results into a more compact, clearer and simpler algorithm that can be utilized to solve also the planar case and can be seen as a formal interpretation of GT's sketch. Indeed, Section \ref{sec:min_projective} unifies, in a sense, HS's algorithm and GT's sketch. Put differently, solving the minimization of $D$ on a tree under planarity is equivalent to solving the projective case for a tree rooted at a specific vertex. For instance, the minimum $D$ under planarity for the tree in Fig. \ref{fig:OPl:first_proj_gt_pla} is obtained when calculating the minimum $D$ under projectivity when the tree is rooted at the vertex marked with a square in Fig. \ref{fig:OPl:first_proj_gt_pla}(b). Section \ref{sec:conclusions} draws some general conclusions and indicates some future paths for research.

\section{Notation and review}
\label{sec:notation_review}

Throughout this paper we use $\Ftree=(V,E)$ to denote a free tree, and $\Rtree=(V,E;\Root)$ to denote a tree $T$ rooted at a vertex $\Root$ where $n=|V|$. Free trees have undirected edges, and rooted trees have directed edges; we consider the edges of a rooted tree to be oriented away from the root. In rooted trees, we refer to the parent of a vertex $u$ as $p(u)$; in a directed edge $(u,v)$, $p(v)=u$. We use $\SubRtree{u}$ to denote a subtree of $\Rtree$ rooted at $u\in V$ (if $u=\Root$ then $\SubRtree{u}=\Rtree$), and $\neighs{u}$ to denote the set of neighbors of vertex $u$ in $T$. We call $\SubRtree{v}$ an {\em immediate subtree} of $\SubRtree{u}$ rooted at $v$ if $(u,v)\in E(\Rtree)$. Extending the notation in \cite{Hochberg2003a}, we use $\SortedSubRtree{1}[u], \cdots, \SortedSubRtree{k}[u]$ to denote the $k$ immediate subtrees of a subtree $\SubRtree{u}$ of $\Rtree$ sorted decreasingly by size. We also use $n_1\ge\cdots\ge n_k\ge 1$ to denote their sizes, i.e., $n_{i}$ denotes the size of $\SortedSubRtree{i}[u]$; we omit the vertex when referring to immediate subtrees of $\Rtree$. Henceforth assume, without loss of generality, that $k$ is even. Recall that $\arr(u)$ is the position of $u\in V$ in the linear arrangement.

\begin{figure}
	\centering
	\includegraphics[scale=0.85]{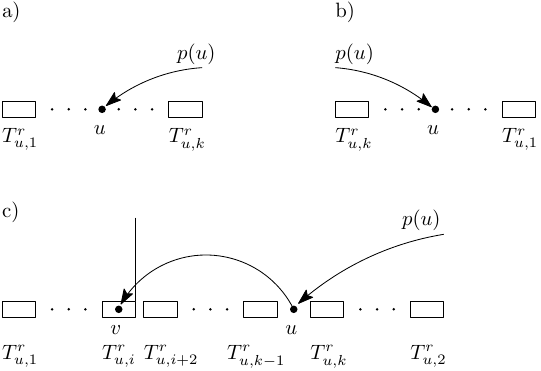}
	\caption{a,b) Optimal arrangements of $\SubRtree{v}$ according to the relative position of $v$ with respect to $v$'s parent. c) Depiction of the directed edges $(p(u),u),(u,v)\in E$ in an optimal projective arrangement, divided into the anchor (the part of the edge $(u,v)$ to the left of the vertical line), and the coanchor (the part of the edge $(u,v)$ to the right). The length of the anchor of edge $(p(u),u)$ is the sum $n_j$ for even $j\in[2,k]$.}
	\label{fig:OPr:embedding_subtrees}
\end{figure}

Now we summarize the core ideas and tools derived by HS \cite{Hochberg2003a}. Firstly, using Lemmas 6, 11 in \cite{Hochberg2003a}, it is easy to see that an optimal projective arrangement of $\Rtree$ is obtained by arranging the immediate subtrees of $\Rtree$ inwards, decreasingly by size and on alternating sides, namely $\SortedSubRtree{1}, \SortedSubRtree{3}, \cdots, \Root, \cdots, \SortedSubRtree{4}, \SortedSubRtree{2}$ or $\SortedSubRtree{2}, \SortedSubRtree{4}, \cdots, \Root, \cdots, \SortedSubRtree{3}, \SortedSubRtree{1}$. Immediate subtrees of $\Rtree$ can be arranged in any of the two orders, whereas immediate subtrees of $\SubRtree{u}$, $u\neq\Root$ have to be placed according to the side in which $u$ is placed with respect to $p(u)$: if $u$ is placed to $p(u)$'s left then the optimal order is $\SortedSubRtree{1}[u], \SortedSubRtree{3}[u], \cdots, u, \cdots, \SortedSubRtree{4}[u], \SortedSubRtree{2}[u]$ (Fig. \ref{fig:OPr:embedding_subtrees}(a)), and if $u$ is placed to $p(u)$'s right the optimal order is $\SortedSubRtree{2}[u], \SortedSubRtree{4}[u], \cdots, u, \cdots, \SortedSubRtree{3}[u], \SortedSubRtree{1}[u]$ (Fig. \ref{fig:OPr:embedding_subtrees}(b)). Notice that the root is not covered in any of these planar arrangements, as required by the projectivity constraint \cite{Kuhlmann2006a,Melcuk1988a}.

\begin{figure}
	\centering
	\includegraphics[scale=0.95]{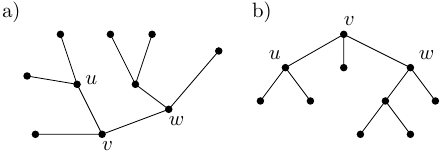}
	\caption{a) A free tree with $s(u,v)=7$, $s(v,u)=3$ and $s(v,w)=s(w,v)=5$. b) The free tree in a) rooted at $v$; $|V(\SubRtree{u}[v])|=s(v,u)$. Borrowed from \cite[Fig. 7]{Hochberg2003a}.}
	\label{fig:OPr:directional_sizes}
\end{figure}

Secondly \cite[Theorem 12]{Hochberg2003a}, an optimal planar arrangement of a free tree $\Ftree$ is obtained when $\Ftree$ is rooted at one of its centroidal vertices. Therefore, an optimal planar arrangement of a free tree $\Ftree$ is an optimal projective arrangement of $\Rtree[c]$, where $c$ denotes one of the (possible two) centroidal vertices of $\Ftree$. For the calculation of a centroidal vertex, HS defined $s(u,v)$, which we call {\em directional} size of subtrees. The directional size $s(u,v)$ in a free tree $\Ftree$, for $\{u,v\}\in E(\Ftree)$, is the size of $\SubRtree{v}[u]$ (Fig.~\ref{fig:OPr:directional_sizes}). Notice that $s(v,u)+s(u,v)=n$. They also outlined a way of calculating all of the $s(u,v)$ in $\bigO{n}$ time \cite[Section 6]{Hochberg2003a}, but did not provide any pseudocode; here we provide it in Algorithm \ref{algo:compute_suvs:free_tree}. Using the $s(u,v)$ for all edges in $T$, we can construct a sorted adjacency list of the tree which we denote as $L$, with the pseudocode given in Algorithm \ref{algo:compute_sorted_L:free_tree}, and with it we calculate one of the centroidal vertices. Algorithm \ref{algo:compute_centroid} reports the pseudocode for the calculation of the centroidal vertex. All algorithms have $\bigO{n}$-time and $\bigO{n}$-space complexity.

We also need to consider the rooting of the list $L$ with respect to a given vertex $w$, denoted as $L^w$. This operation is called $\textsc{root\_list}(L,w)$ in the pseudocode. It transforms the representation of an undirected tree into a directed tree and consists of the action of removing edges of the form $(u,p(u))$, where $u\neq w$, from $L$, starting at the given vertex $w$ which acts as a root. In other words, vertex $w$ induces an orientation of the edges towards the leaves (i.e., away from $w$), and we have to remove one of the two edges $(u,v),(v,u)$ from $L$ for every $\{u,v\}\in E$. Since this can be done fairly easily in linear time, we do not give the pseudocode for this operation.

\begin{algorithm}
	\caption{Calculation of directional sizes for free trees. Cost $\bigO{n}$ time, $\bigO{n}$ space.}
	\label{algo:compute_suvs:free_tree}
	\DontPrintSemicolon
	
	\SetKwProg{Fn}{Function}{ is}{end}
	\Fn{\textsc{compute\_s\_ft}$(\Ftree)$} {
		\KwInTwo{$\Ftree$ free tree.}
		\KwOutTwo{$S=\{(u,v,s(u,v)),(v,u,s(v,u)) \;|\; \{u,v\}\in E\}$.}
	
		$S \gets \emptyset$ \;
		$u_* \gets $ choose an arbitrary vertex \;
		\For {$v\in \neighs{u_*}$} {
			$(\_,S') \gets $\textsc{comp\_s\_ft\_rec}($\Ftree, (u_*,v)$) \;
			$S \gets S \cup S'$
		}
		\Return $S$
	}
	
	\SetKwProg{Fn}{Function}{ is}{end}
	\Fn{\textsc{comp\_s\_ft\_rec}$(\Ftree, (u,v))$} {
		\KwInTwo{$\Ftree$ free tree, $(u,v)$ directing edge.}
		\KwOutTwo{$s$ the size of $\SubRtree{v}[u_*]$ in vertices, $S = \{(u,v, s(u,v)), (v,u, s(v,u)) \;|\; \{u,v\}\in E(\SubRtree{v}[u_*])\}$.}
	
		$s\gets 1$ \;
		\For {$w\in \neighs{v}$} {
			\If {$w\neq u$} {
				$(s', S') \gets$ \textsc{comp\_s\_ft\_rec}($\Ftree, (v,w)$) \;
				$s \gets s + s'$ \;
				$S \gets S \cup S'$ \;
			}
		}
		\tcp{$s=s(u,v)$, $n-s=s(v,u)$}
		\tcp{Append at end in $\bigO{1}$}
		$S \gets S \cup \{(u,v, s), (v,u, n-s)\}$ \;
		\Return $(s,S)$
	}
\end{algorithm}

\begin{algorithm}
	\caption{Calculation of the sorted adjacency list for free trees. Cost $\bigO{n}$ time, $\bigO{n}$ space.}
	\label{algo:compute_sorted_L:free_tree}
	\DontPrintSemicolon
	
	\SetKwProg{Fn}{Function}{ is}{end}
	\Fn{\textsc{sorted\_adjacency\_list\_ft}$(\Ftree)$} {
	\KwInTwo{$\Ftree$ free tree.}
	\KwOutTwo{$L$, the decreasingly-sorted adjacency list of $\Ftree$.}
	
		\tcp{Algorithm \ref{algo:compute_suvs:free_tree}}
		$S\gets \textsc{compute\_s\_ft}(\Ftree)$ \;
		Sort the triples $(u,v,s)$ in $S$ decreasingly by $s$ using counting sort \cite{Cormen2001a} \label{algo:sorting_suvs:free_tree}\;
		$L \gets \{\emptyset\}^n$ \;
		\For {$(u,v,s)\in S$} {
			\tcp{Append at end in $\bigO{1}$}
			$L[u] \gets L[u] \cup {(v,s)}$ \;
		}
		\Return $L$
	}
\end{algorithm}

\begin{algorithm}
	\caption{Calculation of a centroidal vertex of a free tree. Cost $\bigO{n}$ time, $\bigO{n}$ space.}
	\label{algo:compute_centroid}
	\DontPrintSemicolon
	
	\SetKwProg{Fn}{Function}{ is}{end}
	\Fn{\textsc{find\_centroidal\_vertex}$(\Ftree)$} {
	\KwInTwo{$\Ftree$ free tree.}
	\KwOutTwo{A centroidal vertex of $\Ftree$.}
	
		\tcp{Algorithm \ref{algo:compute_sorted_L:free_tree}}
		$L \gets \textsc{sorted\_adjacency\_list\_ft}(\Ftree)$ \;
		\Return \textsc{find\_centroidal\_vertex$(\Ftree, L)$}
	}
	
	\SetKwProg{Fn}{Function}{ is}{end}
	\Fn{\textsc{find\_centroidal\_vertex}$(\Ftree, L)$} {
	\KwInTwo{$\Ftree$ free tree, $L$ sorted adjacency list of $\Ftree$.}
	\KwOutTwo{A centroidal vertex of $\Ftree$.}
		$u\gets $ choose an arbitrary vertex \;
		\While {true} {
			\tcp{$\bigO{1}$ time since $L[u]$ is sorted}
			$(v,s) \gets$ largest entry in $L[u]$ \;
			\lIf {$s > n/2$} {
				$u\gets v$
			}
			\lElse {
				\Return $u$
			}
		}
	}
\end{algorithm}
\section{Minimum planar linear arrangements}
\label{sec:min_planar}


In \cite{Hochberg2003a}, HS present an easy-to-understand algorithm to calculate a minimum planar linear arrangement for any free tree in linear time. The idea behind the algorithm was presented in Section \ref{sec:notation_review}. The implementation has two procedures, {\tt embed} and {\tt embed\_branch}, that perform a series of actions in the following order:
\begin{itemize}
\item Procedure {\tt embed} gets one centroidal vertex, $c$, uses it as a root and orders its immediate subtrees by size.

\item Procedure {\tt embed} puts immediate subtrees with an even index in one side of the arrangement and immediate subtrees with odd index in the other side (the bigger the subtree, the farther away from the centroidal vertex), calling procedure {\tt embed\_branch} for every subtree.

\item Procedure {\tt embed\_branch} calculates recursively a displacement of all nodes with respect to the placement of the centroidal vertex (of the whole tree) in the linear arrangement. 

\item Procedure {\tt embed} calculates the centroidal vertex's position (the sum of sizes of trees on the left of the centroidal vertex) and applies the displacement to the rest of the nodes.
\end{itemize}

\begin{algorithm}
	\caption{Step (5) from procedure \textsc{embed}.}
	\label{algo:embed_branch-embed-5}
	\DontPrintSemicolon
	$\pi(c)\gets leftSum + 1$ \;
	$relPos[c]\gets 0$ \;
	\For {each vertex $v$} {
		$\pi(v)\gets \pi(c) + relPos[v]$ \;
	}
\end{algorithm}

From Algorithm \ref{algo:embed_branch-embed-5}, we can see that vector {\tt relPos} must contain the displacement of all nodes from the position of the centroidal vertex in the linear arrangement. Note that these are only the last lines of {\tt embed}. The problem lays in procedure {\tt embed\_branch}, which does not calculate correctly the displacement vector {\tt relPos}. In Algorithm \ref{algo:embed_branch}, we give a correct version of procedure {\tt embed\_branch}, where changes with respect to HS's version are marked in red. Lines \ref{algo:embed_branch:first} to \ref{algo:embed_branch:add_underanchor_base} are needed to calculate the correct displacement. For a vertex $u\neq c$, variable {\em under\_anchor} is the number of nodes of $\SubRtree{u}[c]$ between $u$ and $p(u)$. Adding {\em under\_anchor} to parameter {\em base} (line \ref{algo:embed_branch:add_underanchor_base}), we obtain the correct displacement. There is also a slight modification in the recursive calls (lines \ref{algo:embed_branch:recursive_call:even} and \ref{algo:embed_branch:recursive_call:odd}) which is the addition of all the parameters needed.

\begin{algorithm}
	\caption{\textsc{embed\_branch} corrected}
	\label{algo:embed_branch}
	\DontPrintSemicolon
	
	\SetKwProg{Fn}{Function}{ is}{end}
	\Fn{\textsc{embed\_branch}$(L^c, v, base, dir, relPos)$} {
	\KwInTwo{(Rooted) sorted adjacency list $L^c$ for $\Rtree[c]$ as described in Section \ref{sec:notation_review}; $v$ the root of the subtree to be arranged; $base$ the displacement for the starting position of the subtree arrangement; $dir$ whether or not $v$ is to the left or to the right of its parent.}
	\KwOutTwo{$relPos$ contains the displacement from the centroidal vertex of all nodes of the subtree.}
	
		\tcp{the children of $v$ decreasingly sorted by size}
		$C_v \gets L^c[v]$ \;
		$before\gets after\gets 0$ \;
		
		\begingroup
		\color{red}
		$under\_anchor\gets 0$ \label{algo:embed_branch:first} \;
		\For {$i = 2$ \bf{to} $\vert C_v\vert$ \bf{step} $2$}
		{
			\tcp{$v$'s $i$-th child, $|V(\SubRtree{v_i}[c])|$ its size}
			$v_i, \Nvert{i} \gets C_v[i]$ \;
			$under\_anchor\gets under\_anchor + \Nvert{i}$ \; 
		}
		$base\gets base + dir*(under\_anchor + 1)$ \label{algo:embed_branch:add_underanchor_base} \;
		\endgroup
		
		\For {$i = $ $\vert C_v\vert$ \bf{downto} $1$} {
			$v_i, \Nvert{i} \gets C_v[i]$\;
			\If {$i$ is even} {
				\begin{tabular}{@{\hspace*{0.0em}}l@{}}
					\textsc{embed\_branch}$(L^c, v_i$, \\
					$\quad base - dir*before, -dir, relPos)$
				\end{tabular} \label{algo:embed_branch:recursive_call:even} \;
				
				$before\gets before + n_i$ \;
			}
			\Else {
				\begin{tabular}{@{\hspace*{0.0em}}l@{}}
					\textsc{embed\_branch}$(L^c, v_i$, \\
					$\quad base + dir*after, dir, relPos)$
				\end{tabular} \label{algo:embed_branch:recursive_call:odd} \;
				$after\gets after + n_i$ \;
			}
		}
		$relPos[v]\gets \textcolor{red}{base} $\;
	}
\end{algorithm}

We should note that {\tt embed} needs to calculate a sorted adjacency list $L$ to calculate a centroidal vertex $c$ for $\Rtree[c]$ (Algorithm \ref{algo:compute_centroid}). However, in order to calculate the arrangement, we need $L$ to be rooted at $c$, then we use $L^c$ (see Section \ref{sec:notation_review}, explanation of \textsc{root\_list}).

In Section \ref{sec:min_projective}, we give an even simpler algorithm that can be seen as a different interpretation of HS's algorithm as it uses the same idea for ordering the subtrees but instead of calculating displacements for nodes it only uses the interval of positions where a subtree must be arranged.  

Prior to HS's work, Iordanskii \cite{Iordanskii1987a} presented an algorithm to solve the task of minimizing $D$ under the planarity constraint. He devised a different approach to solve the same problem: given a free tree, the algorithm roots the tree at its centroid, and then separates the tree into chains of vertices, which have to be arranged in such a way that a planar arrangement is produced. The proper selection of the chains, coupled with the proper labeling of their vertices, produces a minimum planar arrangement. An outline of the algorithm that is applied on $\Rtree[c]$ is as follows \cite{Iordanskii2014a}:
\begin{enumerate}
\item Select an arbitrary vertex $v_0$ in the current decomposition subtree (initial tree).
\item Go from vertex $v_0$ along the branches with the greatest number of vertices to some hanging vertex $v_i$.
\item Starting from vertex $v_i$, construct a chain along the branches with the largest number of vertices to some other hanging vertex $v_j$.
\item Assign the highest and lowest numbers to the vertices $v_i$ and $v_j$ from the range allocated for the current decomposition subtree ($1$ and $n$ for the initial tree).
\item Enumerate monotonically the chain connecting the vertices $v_i$ and $v_j$, leaving the corresponding ranges of numbers for each selected decomposition subtree.
\item The procedure recursively repeats until all vertices are numbered.
\end{enumerate}
The algorithm requires $\bigO{n}$ comparison operations and $\bigO{n\log{n}}$ additional memory.

Iordanskii's approach differs markedly from HS's algorithm, e.g., using chains instead of anchors, and here we have focused on deriving a couple of algorithms for the projective case that stems from HS's algorithm for the planar case.

\section{Minimum projective linear arrangements}
\label{sec:min_projective}

The two algorithms for the projective that are presented in this section have $\bigO{n}$-time and $\bigO{n}$-space complexity, hence our upper bound for the projective case is tighter than that given by PL \cite{Park2009a}. The first algorithm is derived from HS's for the planar case (Algorithm \ref{algo:HS_adaptation}). This algorithm is obtained after extracting the relevant part from HS's original algorithm, adapting it and simplifying procedure {\tt embed}. The simplifications have to do with reducing the computations that Algorithm \ref{algo:compute_suvs:free_tree} does, which are not necessary in the projective variant (Algorithms \ref{algo:compute_suvs:rooted_tree} and \ref{algo:compute_sorted_L:rooted_tree}). Algorithm \ref{algo:compute_suvs:rooted_tree} is the simplified version of \ref{algo:compute_suvs:free_tree} that calculates only the sizes of the subtrees $\SubRtree{u}$ of $\Rtree$ for every vertex $u$ of $\Rtree$; Algorithm \ref{algo:compute_sorted_L:rooted_tree} constructs the rooted sorted adjacency list of a rooted tree $\Rtree$ with less calculations than Algorithm \ref{algo:compute_sorted_L:free_tree}. There is no equivalent to Algorithm \ref{algo:compute_centroid} for rooted trees because we do not need to look for any centroidal vertex. Finally, one has to use the correction of the subprocedure {\tt embed\_branch} Algorithm \ref{algo:embed_branch}. Algorithm \ref{algo:HS_adaptation} inherits the $\bigO{n}$-time and $\bigO{n}$-space complexity from HS's algorithm.

\begin{algorithm}
	\caption{Calculation of size of subtrees for rooted trees. Cost $\bigO{n}$ time, $\bigO{n}$ space.}
	\label{algo:compute_suvs:rooted_tree}
	\DontPrintSemicolon
	
	\SetKwProg{Fn}{Function}{ is}{end}
	\Fn{\textsc{compute\_s\_rt}$(\Rtree)$} {
		\KwInTwo{$\Rtree$ rooted tree.}
		\KwOutTwo{$S = \{(u,v, s(u,v)) \;|\; (u,v)\in E\}$.}
	
		$S \gets \emptyset$ \;
		\For {$v\in \neighs{\Root}$} {
			$(\_,S') \gets \textsc{comp\_s\_rt\_rec}(\Rtree, (\Root,v))$ \;
			$S \gets S \cup S'$
		}
		\Return $S$
	}
	
	\SetKwProg{Fn}{Function}{ is}{end}
	\Fn{\textsc{comp\_s\_rt\_rec}$(\Rtree, (u,v))$} {
		\KwInTwo{$\Rtree$ rooted tree, $(u,v)$ directing edge.}
		\KwOutTwo{$s$ the size of $\SubRtree{v}$ in vertices, $S = \{(u,v, s(u,v)) \;|\; (u,v)\in E(\SubRtree{v})\}$.}
	
		$s\gets 1$ \;
		\tcp{Iterate on the out-neighbours of $v$}
		\For {$w\in \neighs{v}$} {
			$(s', S') \gets \textsc{comp\_s\_rt\_rec}(\Rtree, (v,w))$ \;
			$s \gets s + s'$ \;
			$S \gets S \cup S'$ \;
		}
		\tcp{$s=s(u,v)$}
		\tcp{Append at end in $\bigO{1}$}
		$S \gets S \cup \{(u,v, s)\}$ \;
		\Return $(s,S)$
	}
\end{algorithm}

\begin{algorithm}
	\caption{Calculation of the sorted adjacency list for rooted trees. Cost $\bigO{n}$ time, $\bigO{n}$ space.}
	\label{algo:compute_sorted_L:rooted_tree}
	\DontPrintSemicolon
	
	\SetKwProg{Fn}{Function}{ is}{end}
	\Fn{\textsc{sorted\_adjacency\_list\_rt}$(\Rtree)$} {
	\KwInTwo{$\Rtree$ rooted tree.}
	\KwOutTwo{$L$ the decreasingly-sorted adjacency list of $\Rtree$.}
	
		\tcp{Algorithm \ref{algo:compute_suvs:rooted_tree}}
		$S\gets \textsc{compute\_s\_rt}(\Ftree)$ \;
		Sort the triples $(u,v,s)$ in $S$ decreasingly by $s$ using counting sort \cite{Cormen2001a} \label{algo:sorting_suvs:rooted_tree}\;
		$L \gets \{\emptyset\}^n$ \;
		\For {$(u,v,s)\in S$} {
			\tcp{Append at end in $\bigO{1}$}
			$L[u] \gets L[u] \cup {(v,s)}$ \;
		}
		\Return $L$
	}
\end{algorithm}

\begin{algorithm}
	\caption{Adaptation of HS's main procedure for the projective case.}
	\label{algo:HS_adaptation}
	\DontPrintSemicolon
	
	\SetKwProg{Fn}{Function}{ is}{end}
	\Fn{\textsc{HS\_Projective}$(\Rtree)$} {
	\KwInTwo{$\Rtree$ rooted tree at $\Root$.}
	\KwOutTwo{An optimal projective arrangement $\arr$.}
		
		\tcp{Steps 1 and 3 of HS's algorithm}
		\tcp{Algorithm \ref{algo:compute_sorted_L:rooted_tree}}
		$L^\Root \gets \textsc{sorted\_adjacency\_list\_rt}(\Rtree)$ \;
		$relPos \gets \{0\}^n$ \;
		
		$leftSum \gets rightSum \gets 0$ \;
		\For {$i=k$ \textbf{downto} $1$} {
			\If {$i$ is even} {
				\tcp{Algorithm \ref{algo:embed_branch}}
				\begin{tabular}{@{\hspace*{0.0em}}l@{}}
					\textsc{embed\_branch}$(L^\Root, v_i,rightSum,1,$ \\
					$\quad relPos)$
				\end{tabular}
				$rightSum \gets rightSum + n_i$ \;
			}
			\Else {
				\tcp{Algorithm \ref{algo:embed_branch}}
				\begin{tabular}{@{\hspace*{0.0em}}l@{}}
					\textsc{embed\_branch}$(L^\Root, v_i,-leftSum,-1,$ \\
					$\quad relPos)$
				\end{tabular}
				$leftSum \gets leftSum + n_i$ \;
			}
		}
		
		$\arr \gets \{0\}^n$ \tcp{empty arrangement}
		$\arr(\Root) \gets leftSum + 1$\;
		$relPos[\Root] \gets 0$ \;
		\lFor {each vertex $v$} {
			$\arr(v) \gets \arr(\Root) + relPos[v]$ 
		}
		\Return $\arr$
	}
\end{algorithm}

The second algorithm for the projective case is based on a different approach based on intervals (Algorithms \ref{algo:min_projective} and \ref{algo:arrange}). Although the pseudocode given can be regarded as a formal interpretation of GT's sketch \cite{Gildea2007a} its correctness stems largely from the theorems and lemmas given by HS \cite{Hochberg2003a} (summarized in Section \ref{sec:notation_review}). In Algorithm \ref{algo:min_projective} we give the main procedure that includes the call to the embedding recursive procedure, given in Algorithm \ref{algo:arrange}, which could be seen as a combination of HS's methods {\tt embed\_branch} and {\tt embed} excluding the calculation of one of the centroidal vertices \cite{Hochberg2003a}.

Algorithm \ref{algo:arrange} calculates the arrangement of the input tree $\Rtree$ using intervals of integers $[a,b]$, where $1\le a\le b\le n$, that indicate the first and the last position of the vertices of a subtree in the linear arrangement; an approach based on intervals (but using chains) was considered earlier by Iordanskii \cite{Iordanskii2014a}. For the case of $\Rtree$, the interval is obviously $[1,n]$, as seen in the first call to Algorithm \ref{algo:arrange} (line \ref{algo:min_projective:first_rec_call} of Algorithm \ref{algo:min_projective} and line \ref{algo:min_planar:first_rec_call} of \ref{algo:min_planar}). The loop at line \ref{algo:arrange:for_loop} of Algorithm \ref{algo:arrange} is responsible for arranging all immediate subtrees of $\SubRtree{u}$ following the ordering described by HS (Section \ref{sec:notation_review}). Now, let $\SubRtree{u}$ ($u\neq\Root$) be a subtree of $\Rtree$ to be embedded in the interval $[a,b]$, where $u$, $a$ and $b$ are parameters of the recursive procedure. If one of the immediate subtrees of $\SubRtree{u}$, say $\SubRtree{v}$ with $\Nvert{v} = |V(\SubRtree{v})|$, is to be arranged in the available interval farthest to the left of its parent $u$, its interval is $[a, a + \Nvert{v} - 1]$ (lines \ref{algo:arrange:start_big_if_1}-\ref{algo:arrange:end_big_if_1}); when it is to be arranged in the available interval farthest to the right of $u$, its interval is $[b - \Nvert{v} + 1, b]$ (lines \ref{algo:arrange:start_big_else_1}-\ref{algo:arrange:end_big_else_1}). Notice that the side (with respect to $u$) to which subtree $\SubRtree{v}$ has to be arranged is decided by changing the value of the variable {\tt side}, whose initial value is given in either line \ref{algo:arrange:initial_side_right} or line \ref{algo:arrange:initial_side_left} depending on the side to which $u$ has been placed with respect to its parent (said side is given as the parameter $\tau$ to the recursive procedure). After $\SubRtree{v}$ is arranged, we need to update the left and right limits of the arrangement of $\SubRtree{u}$: if the subtree $\SubRtree{v}$ is arranged to the left of $u$, the left limit is to be increased by $\Nvert{v}$ (line \ref{algo:arrange:big_if_2}), and when it is arranged to the right of $u$, the right limit is to be decreased by $\Nvert{v}$ (line \ref{algo:arrange:big_else_2}). When all immediate subtrees of $\SubRtree{u}$ have been arranged (line \ref{algo:arrange:finish_recursive}), only node $u$ needs to be arranged, thus the remaining interval $[a,b]$ has one element, and then $a = b$ and $\pi(u)=a$.

Furthermore, using this recursive procedure, solving the planar variant is straightforward, (Algorithms \ref{algo:min_planar} and \ref{algo:arrange}): given a free tree $\Ftree$, we simply have to find a centroidal  vertex $c$ of $\Ftree$ (Algorithm \ref{algo:compute_centroid}) where to root the tree and then supply $ \Rtree[c]$ and $L^c$ as input of Algorithm \ref{algo:arrange}. This is due to the fact that an optimal  planar arrangement for $\Ftree$ is an optimal projective arrangement for $\Rtree[c]$ \cite{Hochberg2003a}.  Clearly, an optimal planar arrangement for $\Ftree$ needs not be an optimal projective arrangement for $\Rtree$ for $\Root\neq c$, as $\Root$ might be covered. Fig. \ref{fig:OPl:first_proj_gt_pla}(a) shows an optimal projective arrangement of the rooted tree $\Rtree[1]$, which is not an optimal planar arrangement of $\Ftree$; Fig. \ref{fig:OPl:first_proj_gt_pla}(b) shows an arrangement that is both optimal planar for $\Ftree$ and optimal projective for $\Rtree[2]$.

\begin{algorithm}
	\caption{Linear-time calculation of an optimal projective arrangement.}
	\label{algo:min_projective}
	\DontPrintSemicolon
	
	\SetKwProg{Fn}{Function}{ is}{end}
	\Fn{\textsc{arrange\_optimal\_projective}$(\Rtree)$} {
	\KwInTwo{$\Rtree$ rooted tree at $\Root$.}
	\KwOutTwo{An optimal projective arrangement $\arr$.}
		
		\tcp{Algorithm \ref{algo:compute_sorted_L:rooted_tree}}
		$L^\Root \gets \textsc{sorted\_adjacency\_list\_rt}(\Rtree)$ \label{algo:min_projective:adjacency_list} \;
		
		$\arr \gets \{0\}^n$ \tcp{empty arrangement}
		
		\tcp{The starting side `right' is arbitrary.}
		\tcp{Algorithm \ref{algo:arrange}.}
		\textsc{Arrange}$(L^\Root, \Root, \textrm{right}, 1, n, \arr)$ \label{algo:min_projective:first_rec_call} \;
		
		\Return $\arr$
	}
\end{algorithm}

\begin{algorithm}
	\caption{Linear-time calculation of an optimal planar arrangement.}
	\label{algo:min_planar}
	\DontPrintSemicolon
	
	\SetKwProg{Fn}{Function}{ is}{end}
	\Fn{\textsc{arrange\_optimal\_planar}$(\Ftree)$} {
	\KwInTwo{$\Ftree$ free tree.}
	\KwOutTwo{An optimal planar arrangement $\arr$.}
	
		\tcp{Algorithm \ref{algo:compute_sorted_L:free_tree}}
		$L \gets \textsc{sorted\_adjacency\_list\_ft}(\Ftree)$ \;
		
		\tcp{Algorithm \ref{algo:compute_centroid}}
		$c \gets $\textsc{find\_centroidal\_vertex$(\Ftree, L)$} \;
		
		\tcp{list $L$ rooted at $c$ (Section \ref{sec:notation_review})}
		$L^c \gets \textsc{root\_list}(L,c)$ \;
		
		$\arr \gets \{0\}^n$ \tcp{empty arrangement}
		
		\tcp{The starting side `right' is arbitrary.}
		\tcp{Algorithm \ref{algo:arrange}.}
		\textsc{Arrange}$(L^c, c, \textrm{right}, 1, n, \arr)$ \label{algo:min_planar:first_rec_call} \;
		
		\Return $\arr$
	}
\end{algorithm}

\begin{algorithm}
	\caption{Optimal arrangement of a tree according to its sorted adjacency list.}
	\label{algo:arrange}
	\DontPrintSemicolon
	
	\SetKwProg{Fn}{Function}{ is}{end}
	\Fn{\textsc{Arrange}$(L^\Root, u, \tau, a,b, \arr)$} { \label{algo:min_projective:recursive_procedure}
	\KwInTwo{(Rooted) sorted adjacency list $L^\Root$ as described in Section \ref{sec:notation_review}; $u$ the root of the subtree to be arranged; $\tau$ position of $u$ with respect to its parent $p(u)$; $[a,b]$ interval of positions of the arrangement where to embed $\SubRtree{u}$; $\arr$ the partially-constructed arrangement.}
	\KwOutTwo{$\arr$ updated with the optimal projective arrangement for $\SubRtree{u}$ in $[a,b]$.} 
	
		$C_u \gets L^\Root[u]$ \tcp{the children of $u$ decreasingly sorted by size}
		
		\lIf {$\tau$ is $\mathrm{right}$} { side $\gets$ right \label{algo:arrange:initial_side_right} }
		\lElse {							side $\gets$ left  \label{algo:arrange:initial_side_left} }
		
		
		
		\For {$i$ from $1$ to $|C_u|$} { \label{algo:arrange:for_loop}
			\tcp{the $i$-th child of $u$ and its size $\Nvert{v}=|V(\SubRtree{v})|$}
			$v, \Nvert{v} \gets C_u[i]$ \;
			
			\If {$\mathrm{side}$ is $\mathrm{left}$} {
				$\tau_{\mathrm{next}} \gets$ left \label{algo:arrange:start_big_if_1} \;
				$a_{\mathrm{next}}\gets a$ \;
				$b_{\mathrm{next}}\gets a + \Nvert{v} - 1$ \label{algo:arrange:end_big_if_1}
			}
			\Else {
				$\tau_{\mathrm{next}} \gets$ right \label{algo:arrange:start_big_else_1} \;
				$a_{\mathrm{next}}\gets b - \Nvert{v} + 1$ \;
				$b_{\mathrm{next}}\gets b$ \label{algo:arrange:end_big_else_1}
			}
			
			\textsc{Arrange}$(L^\Root, v, \tau_{\mathrm{next}}, a_{\mathrm{next}}, b_{\mathrm{next}}, \arr)$ \label{algo:arrange:rec_call} \;
			
			\lIf {$\mathrm{side}$ is $\mathrm{left}$} { \label{algo:arrange:big_if_2}
				$a \gets a + \Nvert{v}$
			}
			\lElse { \label{algo:arrange:big_else_2}
				$b \gets b - \Nvert{v}$
			}
			side $\gets$ opposite side \;
		}
		$\arr(u) \gets a$ \label{algo:arrange:finish_recursive} \;
	}
\end{algorithm}
 
Algorithm \ref{algo:min_projective}'s time and space complexities are $\bigO{n}$. First, the sorted, and already rooted, adjacency list $L^\Root$ of $\Rtree$ can be computed in $\bigO{n}$ (line \ref{algo:min_projective:adjacency_list}). The running time of Algorithm \ref{algo:arrange} is clearly $\bigO{n}$: the `for' loop (line \ref{algo:arrange:for_loop}) contains constant-time operations, a single recursive call and, since each loop consists of $\degree{u}=|\neighs{u}|$ iterations (for a vertex $u$), the total running time is $\bigO{\sum_{u\in V}\degree{u}}=\bigO{n}$ because every vertex is visited only once. The spatial complexity is $\bigO{n}$: sorting and building the adjacency list $L^u$ requires $\bigO{n}$ space (for any $u$) and Algorithm \ref{algo:arrange} requires extra $\bigO{n}$ space (for the whole stack of the recursion) in the worst case (for path graphs). The same can be said about Algorithm \ref{algo:min_planar}.

\section{Conclusions and future work}
\label{sec:conclusions}

To the best of our knowledge, our work is the first to highlight a relationship between the MLA problem under planarity and the same problem under projectivity. We have shown that HS's algorithm for planarity \cite{Hochberg2003a} contains a subalgorithm to solve the projective case. We suspect that Iordanskii's algorithm for planarity \cite{Iordanskii1987a} may also contain a subalgorithm for the projective case. We have corrected a few aspects of HS's algorithm (Algorithm \ref{algo:embed_branch}). 

We provided two detailed algorithms for the projective case that run without a doubt in $\bigO{n}$ time. One that stems directly from HS's original algorithm for the planar case (Algorithms \ref{algo:HS_adaptation} and \ref{algo:embed_branch}), and another interval-based algorithm (Algorithms \ref{algo:min_projective} and \ref{algo:arrange}) that builds on HS's work but is less straightforward. The latter algorithm leads immediately to a new way to solve the planar case in $\bigO{n}$ time (Algorithms \ref{algo:min_planar} and \ref{algo:arrange}) thanks to the correspondence between the planar case and the projective case that we have uncovered in this article.

GT \cite{Gildea2007a} sketched an algorithm for the projective case and claimed it to run in linear time. PL \cite{Park2009a} added some details but not sufficiently, concluding that it runs in $\bigO{n \log d_{max}}$ time, which, as we have seen, overestimates the actual complexity. During the reviewing process of this paper, it has come to our knowledge a Master Thesis \cite{Bommasani2020a} where an error in GT's algorithm is pointed out. This error does not affect our implementation.

It could be the case that a unified approach for planarity and projectivity could also be adopted for the maximum linear arrangement problem \cite{Hassin2000a}. To the best of our knowledge, a polynomial-time algorithm for the unrestricted case is not forthcoming. An intriguing question is if the maximum linear arrangement problem on trees can be solved in linear time for the projective and planar variants as in the corresponding minimization problem.

\section*{Acknowledgements}

We are grateful to M. Iordanskii for helpful discussions. We thank C. Gómez-Rodríguez for making us aware of reference \cite{Bommasani2020a}. LAP is supported by Secretaria d’Universitats i Recerca de la Generalitat de Catalunya and the Social European Fund. RFC and LAP are supported by the grant TIN2017-89244-R from MINECO (Ministerio de Econom\'{i}a, Industria y Competitividad). RFC is also supported by the recognition 2017SGR-856 (MACDA) from AGAUR (Generalitat de Catalunya). JLE is funded by the grant PID2019-109137GB-C22 from MINECO.

\bibliographystyle{plain}

\end{document}